\documentstyle[epsfig,12pt]{article}

\def\a{\alpha }

\def\L{\Lambda }

\begin{document}
\pagestyle{empty}
\begin{flushright}
CERN-TH.7235/94
\end{flushright}
\hrule width 0pt
\vspace{2.0cm}
\begin{center}
{\bf Andrei L. Kataev
E-mail:KATAEV@CERNVM.CERN.CH};
\\
 Institute for Nuclear Research of the Russian Academy of Sciences,
Moscow 117312, Russia\footnote{On leave of absence at CERN
from February 1994}
\\{\bf     Aleksander V. Sidorov}
\\
Bogoliubov Theoretical Laboratory, Joint Institute for Nuclear Research,
 141980 Dubna, Russia\footnote{E-mail: SIDOROV@THEOR.JINRC.DUBNA.SU}

\end{center}
\vspace{5cm}
\begin{center}
{\bf \Large Abstract}
\end{center}
\vskip 0.1in
We describe the results of our recent work on the determination of
the value of the parameter $\Lambda_{\overline{MS}}^{(4)}$ and of
the $Q^2$-dependence of the Gross--Llewellyn Smith (GLS) sum rule from
the experimental data of the CCFR collaboration on neutrino--nucleon
deep-inelastic scattering, using the Jacobi polynomials QCD analysis.
The obtained results are compared with the information,
available in the literature,
information on the previous experimental measurements of the GLS sum
rule.

\vskip 1cm

\vfill
\begin{flushleft}
CERN-TH.7235/94\\
 May 1994
\end{flushleft}

\noindent{Presented by A.L.K. at the XXIXth Rencontres de Moriond
on QCD and High Energy Hadronic Interactions, M\'eribel, France,
March 1994, and by A.V.S. at the Quarks--94 International
Seminar, Vladimir, Russia, May 1994.}

\thispagestyle{empty}
\vfill\eject
\pagestyle{empty}
\clearpage\mbox{ }\clearpage
\pagestyle{plain}

\newpage
\setcounter{equation} {0}
{\bf 1.}~~~The detailed experimental and theoretical studies of
the deep-inelastic scattering processes  provide the
important information on the applicability of  perturbative
principles for describing the observed $Q^2$-dependence
of the nucleon structure functions (SF) $F_i(x,Q^2)$
($i=2,3$) and of the related  moments
\begin{equation}
M_n(Q^2)=\int_0^1 x^{n-1}F_i(x,Q^2)dx
\label{mom}
\end{equation}
within the framework of QCD. The non-singlet (NS) SF
$xF_3(x,Q^2)=(xF_3^{{\nu}N}+xF_3^{\overline{\nu}N})/2$, which
characterizes the difference of quark and antiquark distributions,
can be measured in the deep-inelastic processes with charged
electroweak currents. The most precise experimental data for this
quantity were recently obtained by the CCFR group at the
Fermilab Tevatron \cite{CCFR}. The comparison of these data
with the perturbative QCD predictions for $xF_3$ was originally
made with the help of the computer program developed in
Ref. \cite{Owens},  based on the solution of the Altarelli--Parisi
equation. The fits were made for various $Q^2$ cuts of the data.
In particular, fitting
the data at $Q^2>10\ GeV^2$, the CCFR collaboration obtained the
following value of  the parameter
$\Lambda_{\overline{MS}}^{(4)}$ \cite{CCFR}:
\begin{equation}
\Lambda_{\overline{MS}}^{(4)}=171\pm 32 (stat) \pm 54 (syst)\ MeV\ .
\label{la}
\end{equation}
This value turns
out to be almost non-sensitive to the variation of the $Q^2$ cuts,
imposed for allowing one to neglect the effects of the high-twist
(HT) contributions at low energies.

Another important result, obtained by the CCFR collaboration, is
the accurate measurement of the Gross--Llewellyn Smith sum rule
\begin{equation}
GLS(Q^2)=\frac{1}{2}\int_0^1
\frac{xF_3^{{\nu}p+\overline{\nu}p}(x,Q^2)}{x}dx
\label{GLS}
\end{equation}
at the scale $Q^2=3\ GeV^2$ \cite{GLSR}:
\begin{equation}
GLS(Q^2=3\ GeV^2)=2.50\pm 0.018 (stat.)\pm 0.078 (syst.).
\label{result}
\end{equation}
It was already shown \cite{ChK} that this value of the GLS sum rule
is consistent with the QCD predictions, provided one takes into account
not only the perturbative QCD corrections \cite{GLSP,LV}
to the quark-parton prediction $GLS_{As}=3$, but the non-perturbative
three-point function QCD sum rules estimates
of the HT contributions \cite{BK} as well. However, the interesting
question of the possibility of extracting
the $Q^2$-dependence of the GLS sum rule from the CCFR data
remained non-studied.

In  recent work \cite{KS} we analysed this problem with the
method of the SF reconstruction over their Mellin
moments, which is based on the following expansion of the SF over the
Jacobi polynomials \cite{Jacobi}-\cite{Jacobi2}:
\begin{equation}
xF_{3}^{N_{max}=12}(x,Q^2)=x^{\alpha}(1-x)^{\beta}
\sum_{n=0}^{N_{max}=12}\Theta_{n}^{\alpha,\beta}(x)
\sum_{j=0}^{n}c_{j}^{(n)}{(\alpha,\beta)}M_{j+2}^{NS}\left(Q^{2}\right),
\label{exp}
\end{equation}
where $c_j^{(n)}(\alpha,\beta)$ are the coefficients that are expressed
through $\Gamma$-functions and the values of the parameters $\alpha$,
$\beta$, namely $\alpha=0.12$ and $\beta=2.0$, were determined in
Ref. \cite{Jacobi1}. The $Q^2$ evolution of the moments $M_{j+2}^{NS}$
can be determined from
the solution of the corresponding renormalization-group
equation expressed in the form  presented in Ref.\cite{Yndurain}.
The basic expansion parameter is of course  the QCD coupling
constant $\alpha_s$, which can be expressed through the QCD
scale parameter $\L_{\overline{MS}}$ in the standard way:
$\a_s(Q^2)/4\pi=1/\beta_0 \ln(Q^2/\L_{\overline{MS}}^2)-
\beta_1 \ln\ln(Q^2/\L_{\overline{MS}}^2)/\beta_0^3
\ln^2(Q^2/\L_{\overline{MS}}^2)$ where
$\beta_0=11-2/3f$, $\beta_1=102-38/3f$.
The relation of Eq. (5), supplemented by the corresponding
solution of the renormalization-group equation for $M_{j+2}^{NS}$,
forms the basis of the computer program created by
the authors of Ref. \cite{Jacobi1}. It was previously tested
and used by the members of the BCDMS collaboration in the course of
a detailed QCD analysis of the experimental data for $F_2(x,Q^2)$
SF of the deep-inelastic muon-nucleon scattering \cite{BCDMS}.
We were using in our studies also this program, thus building  the
bridge between the determination of $\Lambda_{\overline{MS}}^{(4)}$
from $F_2(x,Q^2)$ and $xF_3(x,Q^2)$ SFs.

{\bf 2.}~~~In accordance with the original NS fit of the CCFR
collaboration \cite{CCFR,GLSR}, we have chosen the parametrization of the
parton distributions at fixed momentum transfer in the simplest
form
\begin{equation}
  xF_{3}(x,Q_0^2)=A(Q_0^2)x^{b(Q_0^2)}(1-x)^{c(Q_0^2)} .
\label{e10}
\end{equation}
The constants $A(Q_0^2)$, $b(Q_0^2)$ and  $c(Q_0^2)$ in
Eq. (\ref{e10}) and the QCD scale parameter
$\Lambda_{\overline{MS}}^{(4)}$ were  considered as  free
parameters, which were  determined
for  concrete values of $Q_0^2$.
In order to avoid the influence of the HT effects and the TM
corrections, following the original CCFR analysis
we  used the experimental points of the concrete
CCFR data in the plane
$(x,Q^2)$ with $0.015<x<0.65$ and
$10\ GeV^2<Q^2<501\ GeV^2$. Note that
we  restricted ourselves
by taking $f=4$ throughout the whole work. Moreover, we did
not take into account any threshold
effects in the process of our analysis.

Using Eq. (5) we reconstructed the theoretical
expression for
$xF_3^{N_{max}=12}(x,Q^2,A,b,c,\Lambda)$ in  all
experimental points ($x_{exp}$, $Q^2_{exp}$) (for the detailed
description see Ref. \cite{KS}). The determination
of the free parameters of the fit (namely $A,b,c,\Lambda$)
was made
by minimization of $\chi^2$ by the MINUIT program, which also
automatically
calculated their statistical errors.
The numerical value of the GLS sum rule at different values of the
reference scale
$Q_0^2$ was determined by substituting the concrete values of the
parameters $A(Q_0^2)$, $b(Q_0^2)$ and $c(Q_0^2)$ into Eq. (7) and
calculating its first moment, which determines the expression for
the GLS sum rule. The statistical errors for the sum rule were
calculated from the statystical errors of the parameters
$A(Q_0^2)$, $b(Q_0^2)$ and $c(Q_0^2)$.

The results of the concrete calculations, made for
various $Q_0^2$ points, are presented in
\\Table 1. Figure 1
demonstrates our results for the GLS sum rule obtained in the
process of the next-to-leading order (NLO) fit.
It is worth emphasizing that putting $n$=1 in the NLO expression
for the moments $M_n^{NS}(Q^2)$ one can reconstract the LO expression
for the GLS sum rule only, namely $GLS_{LO}(Q^2)=3[ 1-\a_s/\pi]$.
Threfore, in Fig. 1 we compare our result of the NLO fit with the LO
perturbative expression of the GLS sum rule.

\newpage
\begin{center}
\begin{tabular}{|c|c|c|c|c|c|c|} \hline
\multicolumn{4}{|c|}{NLO} & \multicolumn{3}{c|}{LO}\\ \hline
$ |Q_0^2| $ & $\Lambda_{\overline{MS}}^{(4)}$ & $\chi^2_{d.f.}$ & GLS
& $\Lambda_{{LO}}^{(4)}$ & $\chi^2_{d.f.}$ & GLS  \\
$[GeV^2]$ & $[MeV]$ &
& sum rule   & $[MeV]$  &       & sum rule \\ \hline
  2 &  209 $\pm$ 32  &  71.5/62  & 2.401 $\pm$ 0.126
  & 154 $ \pm $ 16 & 87.6/62&
 2.515 \\
  3 &  213 $\pm$ 31  &  71.5/62  & 2.446 $\pm$ 0.081
  & 154 $ \pm$ 29 & 87.7/62&
 2.525 \\
  5 &  215 $\pm$ 32  &  71.8/62  & 2.496 $\pm$ 0.121
  & 154 $ \pm$ 28 & 88.0/62&
 2.537  \\
  7 &  215 $\pm$ 34  &  72.2/62  & 2.525 $\pm$ 0.105
  & 155 $ \pm$ 27 & 88.3/62&
 2.549  \\
 10 &  215 $\pm$ 35  &  72.6/62  & 2.553 $\pm$ 0.107
 & 154 $ \pm$ 29 & 88.5/62&
 2.558  \\
 15 &  215 $\pm$ 34  &  73.2/62  & 2.583 $\pm$ 0.111
 & 155 $ \pm$ 28 & 88.8/62&
 2.569  \\
 25 &  214 $\pm$ 31  &  74.1/62  & 2.618 $\pm$ 0.113
 & 155 $ \pm$ 17 & 89.2/62&
 2.583  \\
 50 &  213 $\pm$ 33  &  75.4/62  & 2.661 $\pm$ 0.119
 & 155 $ \pm$ 27 & 90.2/62&
 2.603  \\
 70 &  212 $\pm$ 34  &  76.1/62  & 2.680 $\pm$ 0.120
 & 155 $ \pm$ 26 & 90.3/62&
 2.614  \\
 100&  211 $\pm$ 33  &  76.8/62  & 2.699 $\pm$ 0.123
 & 154 $ \pm$ 29 & 90.7/62&
 2.623  \\
 150&  210 $\pm$ 34  &  77.6/62  & 2.720 $\pm$ 0.126
 & 154 $ \pm$ 29 & 91.2/62&
 2.635  \\
 200&  209 $\pm$ 33  &  78.2/62  & 2.735 $\pm$ 0.127
 & 154 $ \pm$ 29 & 91.5/62&
 2.643  \\
 300&  209 $\pm$ 33  &  79.0/62  & 2.755 $\pm$ 0.129
 & 153 $ \pm$ 29 & 92.0/62&
 2.655  \\
 500&  207 $\pm$ 35  &  80.1/62  & 2.779 $\pm$ 0.155
 & 153 $ \pm$ 29 & 92.7/62&
 2.664  \\
\hline
\multicolumn{7}{p{13cm}}{{\bf Table 1.} The results of the LO and NLO QCD fit
of the CCFR $xF_3$ SF data for  $f=4$, $Q^2>10\ GeV^2$, $N_{max}=12$
with the corresponding statistical errors. The symbol
$\chi^2_{d.f.}$ is for the
$\chi^2$ parameter normalized to the number of degrees
 of freedom $d.f.$ }
\\[5mm]
\end{tabular}
\end{center}

\begin{figure} [h]
\begin{center}
\mbox{\epsfig{figure=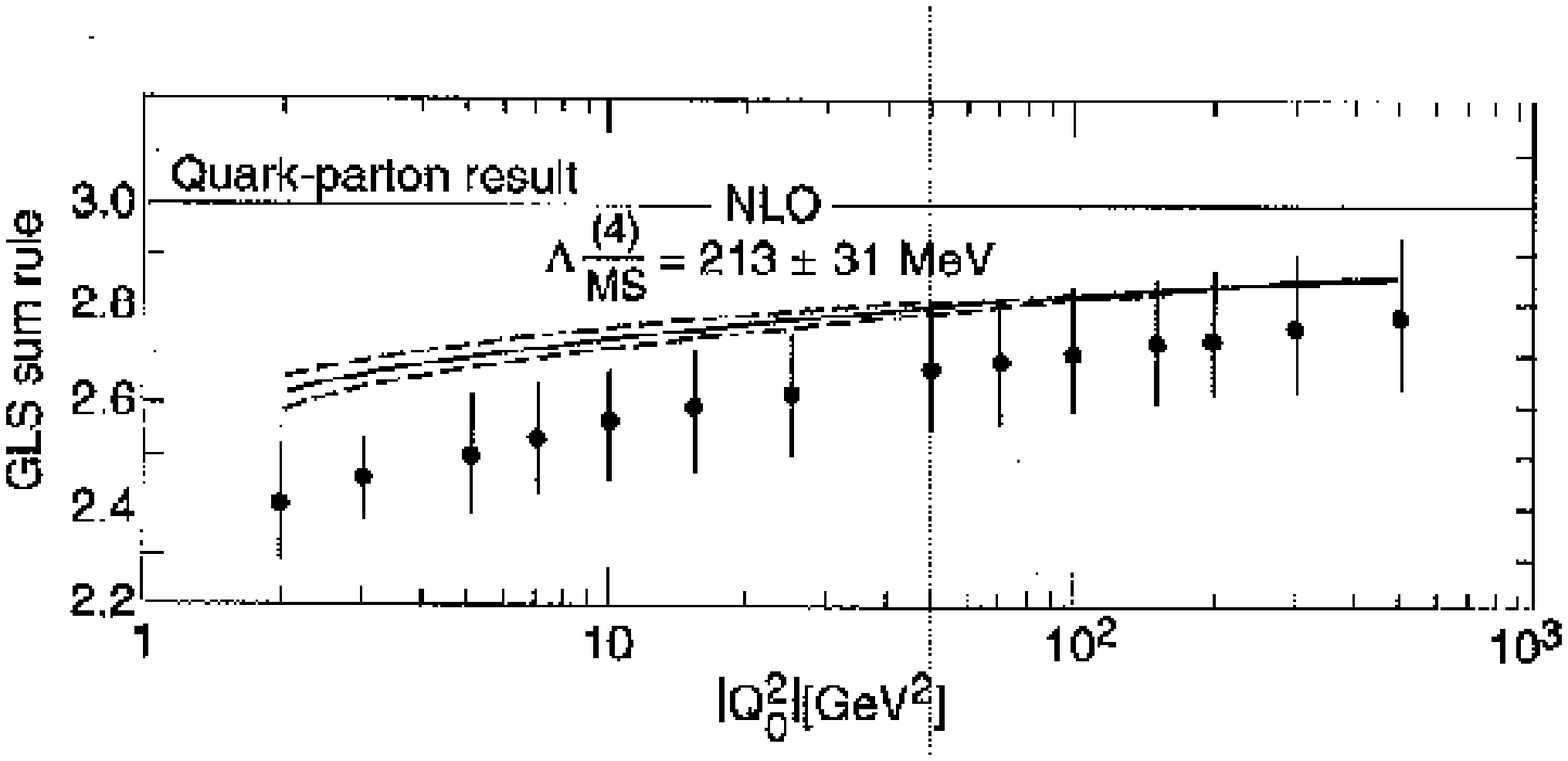,width=12cm}}
\end{center}
{Fig.1. The comparison of the result of the NLO fit
of the $Q^2$ evolution of the GLS sum rule with the statistical
error bars taken into account with the LO perurbative QCD prediction.}
\end{figure}
\newpage

The value of the parameter
$\Lambda_{\overline{MS}}^{(4)}$ in the LO perturbative expression
for the GLS sum rule in Fig. 1 was taken in accordance with the
results of our analysis of the CCFR data for the SF $xF_3$ at the
reference point $Q_0^2=3\ GeV^2$ (see Table 1). Notice, that the
points presented in Fig. 1 are strongly correlated. The explanation
is very simple: they were all obtained from the whole set of data.

{\bf 3.}~~~Taking into account our estimates of the statistical
uncertainties and the estimate,
determined by the CCFR group, of the
systematic uncertainty \cite{GLSR}, we obained  the following value
of the GLS sum rule at the scale $Q_0^2=3\ GeV^2$ \cite{KS}:
\begin{equation}
GLS(Q_0^2=3\ GeV^2)=2.446 \pm 0.081(stat)\pm 0.078 (syst).
\label{glsres}
\end{equation}
This  in the agreement with the result (4) obtained by the
CCFR group. The smaller statistical error of the CCFR result
of Eq. (4) comes from their more refined analysis of this type
of uncertainties. The results of our NLO determination of the
GLS sum rule (see Table 1)  do not contradict either
the previous, less accurate,
measurements of this sum rule at different values of $Q^2$, as
presented in Table 2.

\begin{center}
\begin{tabular}{|c|c|c|c|c|c|c|} \hline
 Collaboration  &  Reference  &  Typical $Q^2$ & Result \\
                &             &   $[GeV^2]$    &        \\  \hline
 CDHS       & \cite{CDHS}  &  1--180  & 3.20 $\pm$ 0.50 \\
 CHARM      & \cite{CHARM} &    10    & 2.56 $\pm$ 0.41 $\pm$ 0.10 \\
 BEBC--     & \cite{BEBC}  &   1--10  & 2.89 $\pm$ 0.33 $\pm$ 0.23 \\
 Gargamelle &              &   10--20 & 3.13 $\pm$ 0.48 $\pm$ 0.28 \\
 CCFRR      & \cite{CCFRR} &    3     & 2.83 $\pm$ 0.15 $\pm$ 0.10 \\
 WA25       & \cite{WA25}  &    3     & 2.70 $\pm$ 0.40 \\
 SKAT       & \cite{SKAT}  &  0.5--10 & 3.10 $\pm$ 0.60 \\
 CCFR       & \cite{CCFR1} &    3     & 2.78 $\pm$ 0.08 $\pm$ 0.13 \\
 CCFR       & \cite{GLSR}  &    3     & 2.50 $\pm$ 0.02 $\pm$ 0.08 \\
 CCFR       & \cite{KS}    &    3     & 2.45 $\pm$ 0.08 $\pm$ 0.08 \\
our analysis&                  &          &                 \\
\hline
\multicolumn{4}{p{12cm}}{{\bf Table 2.} The summary of various
determinations of the GLS sum rule with the corresponding statistical
and systematical uncertainties.}
\\[5mm]
\end{tabular}
\end{center}
Note, that the results of the CHARM collaboration \cite{CHARM}
were obtained after integrating $xF_3$ SF in the interval
$0.0075<x<1$. We expect, that after interpolation of the data in the
region of small $x$ and integrating them in the whole interval
$0<x<1$ the corresponding CHARM results for the GLS sum rule will
approach the quark-parton prediction $GLS_{As}=3$ and will be even
less accurate than the claimed result \cite{CHARM} presented
in Table 2.

The results for $\Lambda_{\overline{MS}}^{(4)}$ obtained from
the NLO fit of the CCFR data, using the Jacobi polynomial expansion
(see Table 1), are in exact agreement with the outcome of the fit
of the BCDMS data for the $F_2(x,Q^2)$ SF with the help of the same
computer program \cite{BCDMS}, namely $\Lambda_{\overline{MS}}^{(4)}
=230 \pm20 (stat)\pm 60(syst)\ MeV$.

The result, used at Fig. 1, namely
$\Lambda_{\overline{MS}}^{(4)}=213\pm31(stat)\ MeV$ ,
which was
 obtained using the expressions for the Mellin moments
$M_{n}^{NS}$, is somewhat larger than the result of Eq. (2) obtained
by the CCFR group with the method based on the solution
of the Altarelli-Parisi equation. A similar feature was previously
observed in the process of the analogous fits of the $xF_3$ less
precise data obtained at Protvino \cite{Protvino}:
the Altarelli-Parisi
method gave $\Lambda_{\overline{MS}}^{(4)}=170\pm 60 (stat)\pm
120(syst)\ MeV$ (compare with Eq. (2)), while the fit over the
Mellin moments resulted in the value $\Lambda_{\overline{MS}}^{(4)}=
230\pm40(stat)\pm 100(syst)\ MeV$, which should be compared with
the results of our fit (see Table 1).
This slight
discrepancy between the central values of the outcomes of different
fits of the same data might be due to the intrinsic features of the two
different ways the perturbative QCD corrections are taken into account.
One can hope that this difference will be minimized
after incorporating  higher-order perturbative QCD effects into both
methods.

{\bf 4.}~~~The $Q^2$ dependence of the GLS sum rule (see Fig. 1),
extracted by us from the CCFR data,
does not contradict the  previous measurements of the $Q^2$ dependence
of this sum rule, made by the BEBC--Gargamelle collaboration
\cite{BEBC} (see Fig. 2) and WA25 collaboration \cite{WA25} (see Fig. 3).

\begin{figure} [h]
\begin{center}
\mbox{\epsfig{figure=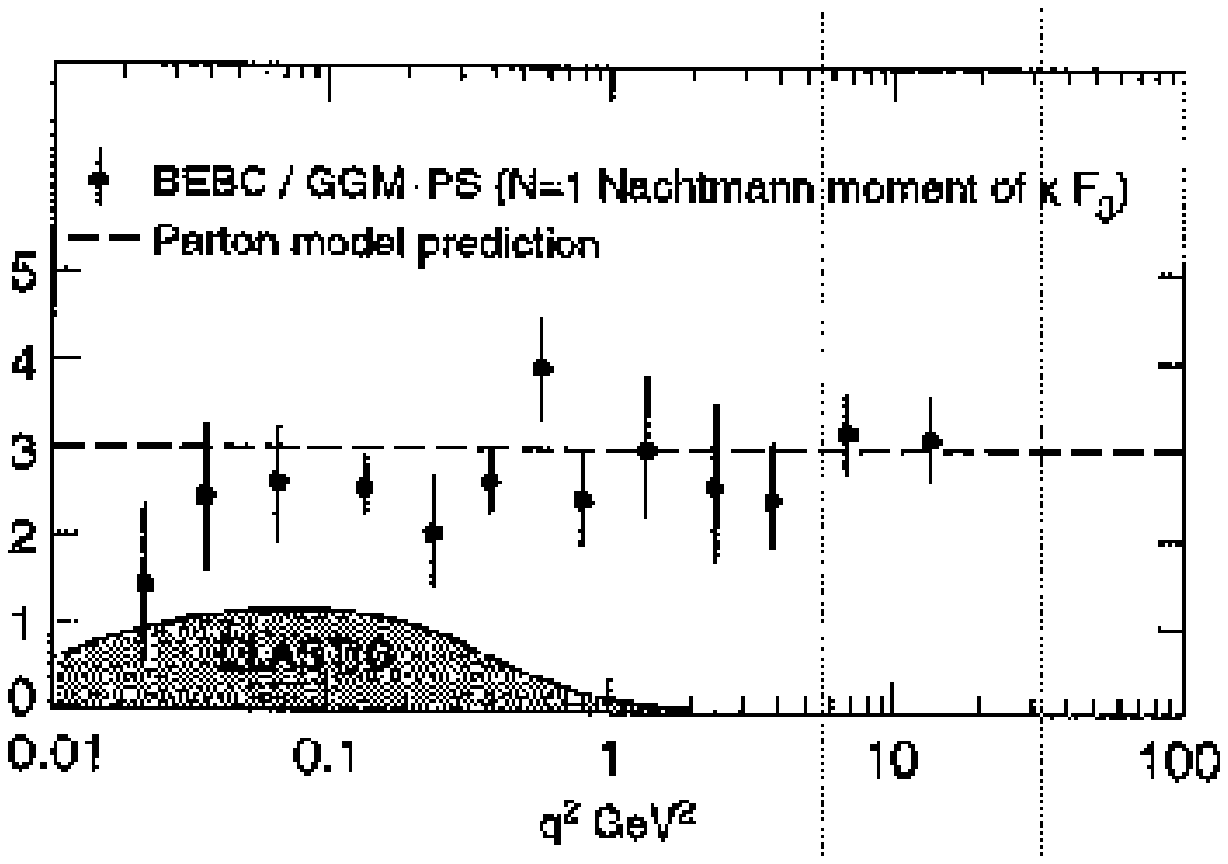,width=10cm}}
\end{center}
{Fig.2. Data on the GLS sum rule from the combined BEBC
narrow-band neon and GGM-PS freon neutrino/antineutrino
experiments \cite{BEBC}. Errors shown are statistical only.}
\end{figure}

\newpage
\begin{figure} [h]
\begin{center}
\mbox{\epsfig{figure=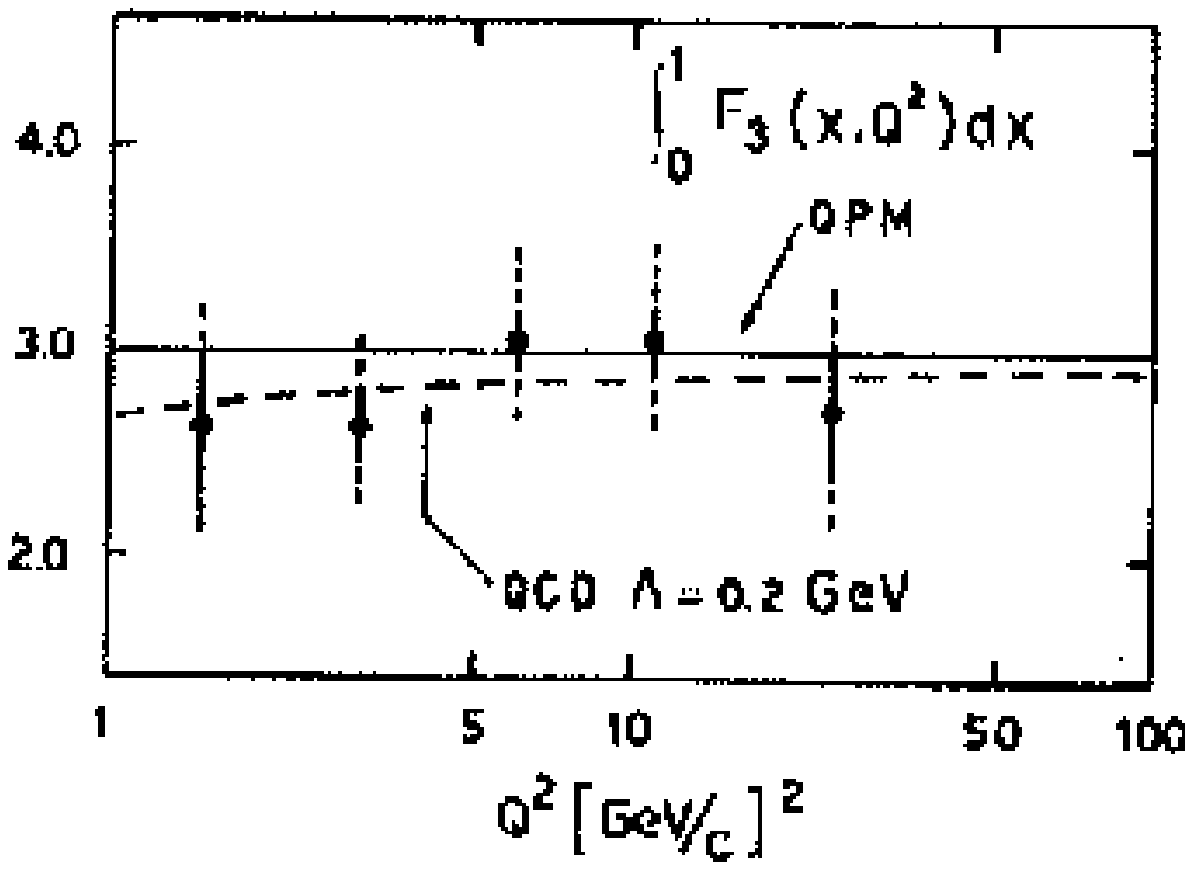,width=10cm}}
\end{center}
{Fig.3. The $Q^2$-dependence of the GLS sum rule, extracted
by WA25 collaboration \cite{WA25}.}
\end{figure}
However, the larger uncertainties of these data
did not allow one to reveal the characteristic behaviour of the
experimental results in the low-energy region, clearly seen in the
analysis of the CCFR data \cite{GLSR,KS}. Indeed, the results of
Eqs. (4) and (10) lie much lower than the
theoretical predictions of the pure
perturbative QCD. This feature demonstrates the importance of
taking  account of the HT contributions in the theoretical
expression for the GLS sum rule. Their general structure is known
from the results of Ref.\cite{SV}.
The corresponding numerical
calculations of these terms were made in Ref. \cite{BK} and more recently
in Ref. \cite{RR}, using the same three-point function QCD sum
rules technique. Combining all available  information about
the GLS sum rule we can write the following theoretical expression:
\begin{equation}
GLS_{QCD}(Q^2)=3\left[1-a-(4.583-0.333f)a^2-(41.441-8.02f+0.177f^2)a^3
-\frac{8}{27}\frac{\langle\langle O\rangle\rangle}{Q^2}\right]
\label{resqcd}
\end{equation}
where $a=\alpha_s/\pi$ is the coupling constant in the
$\overline{MS}$ scheme. The NLO and NNLO perturbative QCD corrections
were calculated in Refs. \cite{GLSP} and \cite{LV} respectively.
Note, that the perturbative expression for the GLS sum rule has
one interesting feature, namely it is related to the perturbative
expression for the $e^+e^-\rightarrow\mbox{hadrons}$
D-function, calculated
at the NLO and NNLO levels in Refs. \cite{ChKT} and \cite{GKL} (see
also \cite{SS}), by the non-trivial generalization of the quark-parton
connection of Ref. \cite{Crewther}. This generalization, discovered
in Ref. \cite{BK}, has the perturbative corrections starting from the
NLO level. They are proportional to the two-loop QCD $\beta$-function.
The theoretical consequences of the appearence of this factor
are not yet
clear. However, even at the current level of understanding it is
possible to conclude, that the results of the work \cite{BK}  provide
the strongest argument in favour of the correctness
of the results of the NLO and NNLO calculations of the GLS sum rule
and the $e^+e^-$-annihilation R-ratio.

\newpage
Let us return to the discussions of the effects of the
HT-contributions in Eq. (12).
The original calculation \cite{BK} of the  matrix element
$\langle\langle O\rangle\rangle$
gave the following estimate $\langle\langle O\rangle\rangle
=0.33\pm0.16\ GeV^2$. The QCD
prediction of Eq. (12) with this value of the HT term was used in
Ref. \cite{ChK} for the extraction of the value of
$\Lambda_{\overline{MS}}^{(4)}$ from the experimental result of
Ref. \cite{GLSR}. We will not discuss here all details of the work
of Ref. \cite{ChK}, but present only final outcomes of the NLO analysis
in the $\overline{MS}$ scheme with HT terms
\begin{equation}
\Lambda_{\overline{MS}}^{(4)}=318\pm23(stat)\pm99(syst)\pm62(twist)
\ MeV
\label{LHT}
\end{equation}
and without HT terms
\begin{equation}
\Lambda_{\overline{MS}}^{(4)}=435\pm20(stat)\pm87(syst)\ MeV\ .
\end{equation}
It can be seen that the HT terms are decreasing the difference of the
extracted values of the parameter $\Lambda_{\overline{MS}}^{(4)}$
from the results of other NLO fits, say from our results of Table 1.
Even better agreement can be obtained after taking into account NNLO
corrections in the GLS sum rule (see Eq. (12)), scheme-dependence
ambiguities (see Ref. \cite{ChK}) or the new estimates of the
HT-contribution, namely $\langle\langle O\rangle\rangle=
0.53\pm0.04\ GeV^2$ \cite{RR}, which
however have surprisingly small error bars of over
 10$\%$ (it is known
that the typical uncertainties of different
three-point  function QCD sum
predictions lie within error bars of at least 30$\%$ ).

Since at the scale $Q^2=3\ GeV^2$ the result of our extraction
of the GLS sum rule value   \cite{KS}
is in agreement with the original result of the CCFR group \cite{GLSR}
(compare Eq. (4) with Eq. (10)), the conclusions of
Ref. \cite{ChK} remain valid in our case also.
Moreover, we consider the
deviation of the $Q^2$ dependence
of the GLS sum rules results that we observed
in the low-energy region from the prediction
of perturbative QCD (see Fig. 1)
as an indication
of the necessity for a  detailed study of the HT effects in the region
of $Q^2<10\ GeV^2$. This conclusion joins the results of the
quantitative analysis \cite{ChK}, \cite{EK} of the effects of
the HT contributions to the GLS sum rule \cite{BK} and
the Bjorken polarized sum rules \cite{BBK} correspondingly
(see also Ref. \cite{RR}) and support
the necessity of taking into considerations of these effects in the
detailed description    of the $Q^2$ dependence of the deep-inelastic
scattering sum rules in the low energy region.

In the high-energy region $Q^2\geq10\ GeV^2$ the
 $Q^2$ behaviour of the GLS sum rule, obtained by us and depicted
in Fig. 1, is in qualitative
agreement with the perturbative QCD expectations. However, at the
quantitative level there are  indications of the existence of
the deviation between theoretical predictions
and the results of our analysis. This phenomenon might
be related to the necessity for detailed studies of the effects of
the NNLO corrections to the NS anomalous dimensions (which are known
at present only for even moments \cite{LRV}) and of the NNLO coefficients
of the NS moments of the $xF_3$ SF \cite{NZ}. Another important task is
an improvement of the understanding of the behaviour of the $xF_3$ SF
in the region of small $x$.
We hope that a possible future analysis
 will allow one to study the $Q^2$ dependence of the GLS sum rule in
more detail and to understand the status of the non-standard theoretical
explanation of the behaviour of the GLS sum rule observed by us at
moderate $Q^2$ \cite{Dorokhov}.

\newpage
{{\bf Acknowledgements}}

We are grateful to  M.H. Shaevitz and W.G. Seligman  for providing us
the CCFR data.
The work of one of us (A.V.S.) is supported by the Russian Fund
for Fundamental Research Grant N 94-02-04548-a.

\end{document}